\documentclass[a4paper, english]{article}
\usepackage[T1]{fontenc}
\usepackage[latin9]{inputenc}
\usepackage{babel}
\usepackage{amsmath,amssymb}
\usepackage{amscd}
\usepackage{amsfonts}
\usepackage{textcomp}
\usepackage{slashed}
\usepackage{cancel}
\usepackage{pb-diagram}
\usepackage{color}
\usepackage[usenames,dvipsnames,svgnames,table]{xcolor}
\usepackage[bookmarks=false,colorlinks=true,linkcolor={red},citecolor=blue,pdfstartview={XYZ null null 1.22},backref,pagebackref]{hyperref}
\usepackage{cleveref}
\renewcommand{\cal}[1]{\mathcal{#1}}
\usepackage{graphicx}
\newcommand{\be}{\begin{equation}}
\newcommand{\ee}{\end{equation}}
\newcommand{\bea}{\begin{eqnarray}}
\newcommand{\eea}{\end{eqnarray}}
\newcommand{\beas}{\begin{eqnarray*}}
\newcommand{\eeas}{\end{eqnarray*}}

\usepackage{soul}

\def\({\left(}  
\def\){\right)}

\title{Relativistic Runge-Lenz vector: from ${\cal N}=4$ SYM to SO(4) scalar field theory}
\begin{document}

\author{J. Alvarez-Jimenez$^1$, I. Cortese$^2$,  J. Antonio Garc\'ia$^3$, \\ D. Guti\'errez-Ruiz$^4$, and J. David Vergara$^5$\\
\small Departamento de F\'isica de Altas Energ\'ias, Instituto de Ciencias Nucleares\\
\small Universidad Nacional Aut\'onoma de M\'exico,\\
\small Apartado Postal 70-543, Ciudad de M\'exico, 04510, M\'exico\\
\small $^1$  javier171188@hotmail.com,  $^2$nachoc@nucleares.unam.mx, $^3$ garcia@nucleares.unam.mx, \\ \small $^4$gutierrezrd90@gmail.com, $^5$vergara@nucleares.unam.mx }

\maketitle


\begin{abstract}

Starting from ${\cal N}=4$  SYM and using an appropriate Higgs mechanism we reconsider the construction of a scalar field theory non-minimally coupled to a Coulomb potential with a relativistic SO(4) symmetry and check for scalar field consistency conditions. This scalar field theory can also be obtained from a relativistic particle Lagrangian with an appropriate implementation of the non-minimal coupling. We provide the generalization of the non-relativistic construction of the Runge-Lenz vector to the relativistic case and show explicitly that this new vector generates the SO(4) algebra. Using the power of the SO(4) symmetry, we calculate the relativistic hydrogen atom spectrum. We provide a generalization of the Kustaanheimo-Stiefel transformation to the relativistic case and relate our results with the corresponding relativistic oscillator. Finally, in the light of these results, we reconsider the calculation of the hydrogen atom spectrum from the cusp anomalous dimension given in \cite{Caron}.

\end{abstract}

\section{Introduction}

Recent developments on integrability of ${\cal N}=4$ SYM opened the way to impressive results in the context of scattering amplitudes. In particular, using a new powerful duality symmetry in \textit{momentum space}, the complete calculation of all tree level amplitudes as well as up to fourth order loops in perturbation theory were accomplished \cite{Bern}. Complemented by results from strong coupling promoted by the AdS/CFT conjecture, gave rise to a new duality between Wilson loops and scattering amplitudes that drastically simplifies the calculation of the corresponding amplitudes. 

In this context, the authors of \cite{Caron} raised the interesting question about the possibility of constructing a consistent relativistic quantum field theory that preserves the analogous of the Runge-Lenz (RL) vector of the Kepler potential in classical mechanics. The surprising answer to this question is that it is indeed possible and that theory is ${\cal N}=4$ SYM processed by a Higgs mechanism that gives mass to some scalars in the field content of that theory. In turn, this interesting result was used to study the relativistic spectrum of the non-BPS two-body bound state of a higgsed ${\cal N}=4$ SYM theory in the limit when one particle mass, the ``proton'' mass, goes to infinity. It is possible to address this problem in the context of large $N$ limit of SYM using integrability techniques and dual conformal symmetry. To find the spectrum we can use a duality relation between the anomalous dimension of the CFT (Wilson loops with a cusp) and the angular momentum, plus a quantization condition \cite{Caron}. This interesting relation was also used to explore the meson spectrum and to compare it with the corresponding spectrum obtained using string theory in the context of AdS/CFT \cite{mateos}. 

The results presented in \cite{Caron} were worked out using numerical methods. Analytic results that confirmed the numerical approach were obtained in \cite{Espindola}. The same two body bound-state spectrum can also be addressed using relativistic quantum mechanics with an enhanced symmetry \cite{Sakata}. 

Our aim in this letter is to construct explicitly the analogous of the RL vector in the relativistic setup and analyze some of its consequences. We present the infinitesimal relativistic transformation generated by this new RL vector and compare the results with the approach followed in \cite{Caron}. Later on, we will also argue that this relativistic RL vector can be used, as in classical mechanics, to make explicit the hidden SO(4) symmetry of the hydrogen atom and to calculate the corresponding relativistic spectrum for this \textit{relativistic hydrogen-like atom}. Our results for this spectrum confirm previous analyses using different approaches \cite{Sakata,Qiang}. Next, we construct a new relativistic Kustaanheimo-Stiefel (KS) duality between different central potentials. Using this duality transformation we show that the relativistic spectrum of the hydrogen-like atom in 3 dimensions is related to the relativistic harmonic oscillator spectrum in 4 dimensions. Our results again confirm previous analyses where the spectrum of the relativistic harmonic oscillator has been constructed from scratch \cite{Qiang2}.   

At first sight, the existence of a relativistic RL vector is a nonsense. It is well known that the orbits of a relativistic particle minimally coupled to a Kepler scalar central potential are open rosettes, and, consequently, the associated symmetry of the classical non-relativistic problem is broken. The relativistic non-degenerated problem possesses less symmetry. So, relativistic effects break the symmetry algebra generated by the angular momentum and the RL vector. That means that the RL vector is not a conserved quantity in the relativistic realm. Nevertheless it is still possible to restore the SO(4) symmetry in the relativistic context if we allow for a non-minimal coupling of the particle with the scalar field. This observation is central to our work. This non-minimal coupling is widely used in the context of a modified Dirac equation with enhanced symmetry algebras in the theory of nuclear spectrum \cite{report}. 

We restrict ourselves to stationary problems, so we choose a time direction in Minkowski spacetime. Additionally, we select a Lorentz frame where the electromagnetic vector potential has the form $A^\mu=(V(r),0,0,0)$, breaking the manifest Lorentz symmetry. Here $r$ is just a spatial coordinate, so the symmetry is reduced from SO(3,1) to SO(3). Then, using the non-minimal coupling, we will enhance our symmetry to SO(4). It is clear that we can not have SO(3,1) and SO(4) in the same description, then our approach keeps only SO(4).  Of course, a good question is if we can have SO(4,1) or, even better, the complete conformal symmetry SO(4,2) by adding more degrees of freedom. Diverse proposals around these problems have been presented in recent literature (see for example \cite{gilmore}).

\section{Reduction from ${\cal N}=4$ Super Yang-Mills to a non-minimally coupled scalar field}

To describe our model we present explicitly the procedure to reduce ${\cal N}=4$   SYM \cite{SYN4, Beisert} to a non-minimally coupled scalar field theory \cite{Sakata}.  Our interest here is to explicitly show the emergence of a scalar field theory with SO(4) symmetry from ${\cal N}=4$ SYM \cite{Brink}  and check the consistency conditions that follow from the equations of motion. This scalar field theory was previously considered in \cite{Sakata}. To undertake this task we consider only the bosonic sector of ${\cal N}=4$ SYM which has the following Lagrangian density 

\begin{equation}\label{SYM-Lag}
{\cal L}=\mathrm{Tr}\left\{ -\frac{1}{4}F^{\mu\nu}F_{\mu\nu}-\frac{1}{2}\sum_{i=1}^{6}D_{\mu}\Phi_{i}D^{\mu}\Phi_{i}+\frac{g^{2}}{4}\sum_{i,j=1}^{6}[\Phi_{i},\Phi_{j}]^{2}\right\} .
\end{equation}

Here the six scalar fields are $N\times N$ traceless hermitian matrices
in the adjoint representation of SU(N). The action of the covariant
derivative on a generic field $W$ is given by
\begin{equation}
D_{\mu}W=\partial_{\mu}W-ig[A_{\mu},W],
\end{equation}
where under a gauge transformation $U$ the matrix gauge field $A_{\mu}$
and the scalar fields $\Phi_{i}$ transform as
\begin{equation}
\Phi_{i}\rightarrow U\Phi_{i}U^{\dagger},\,\,\,\,\,\,\,\,\,\,A_{\mu}\rightarrow UA_{\mu}U^{\dagger}-\frac{i}{g}(\partial_{\mu}U)U^{\dagger}.
\end{equation}
The resulting equations of motion are \cite{Beisert}
\begin{equation}
D_{\mu}F^{\nu\mu}=ig\sum_{i=1}^{6}[\Phi_{i},D^{\nu}\Phi_{i}],\label{eq:eqfmunu}
\end{equation}
\begin{equation}
D_{\mu}D^{\mu}\Phi_{i}=g^{2}\sum_{j=1}^{6}[\Phi_{j,}[\Phi_{j},\Phi_{i}]].\label{eq:eqscalar}
\end{equation}

For simplicity, we choose to work with the group SU(2), following \cite{Schabinger}. Therefore, the fields will be expressible in terms of the Pauli matrices $\tau^a$ as
\begin{equation}\label{eq:decomp}
\Phi_{i}=\Phi_{i}^{a}\frac{\tau^{a}}{2}=\frac{1}{2}\begin{pmatrix}\Phi_{i}^{0} & \Phi_{i}^{-}\\
\Phi_{i}^{+} & -\Phi_{i}^{0}
\end{pmatrix},\,\,\,\,\,\,\,A_{\mu}=A_{\mu}^{a}\frac{\tau^{a}}{2}=\frac{1}{2}\begin{pmatrix}A_{\mu}^{3} & A_{\mu}^{1}-iA_{\mu}^{2}\\
A_{\mu}^{1}+iA_{\mu}^{2} & -A_{\mu}^{3}
\end{pmatrix},
\end{equation}
where $\Phi_{i}^{\pm}=\Phi_{i}^{1}\pm i\Phi_{i}^{2}$.
We now introduce the spontaneous symmetry breaking by giving a vacuum
expectation value $v$ to $\Phi_{1}$ \cite{Sakata,Schabinger,Alday}
\begin{equation}\label{vevforPhi1}
\Phi_{1}=\frac{1}{2}\begin{pmatrix}\Phi_{1}^{0}+v & 0\\
0 & -\Phi_{1}^{0}-v
\end{pmatrix},
\end{equation}
and taking the other fields as
\begin{equation}
\Phi_{2}=\frac{1}{2}\begin{pmatrix}0 & \Phi_{2}^{-}\\
\Phi_{2}^{\text{+}} & 0
\end{pmatrix},\,\,\,\,\,\Phi_{i}=0,\,\,i=3,4,5,6;\,\,\,\,\,\,\,A_{\mu}=A_{\mu}^{a}\frac{\tau^{a}}{2}=\frac{1}{2}\begin{pmatrix}A_{\mu}^{3} & 0\\
0 & -A_{\mu}^{3}
\end{pmatrix}.
\end{equation} 
We switched off the charged components $A^1_\mu,A^2_\mu$ of the connection field $A_\mu$. 

Now the Lagrangian (\ref{SYM-Lag}) reads 
\begin{equation}\nonumber
{\cal L}= -\frac{1}{8}F^{\mu\nu}F_{\mu\nu}
-\frac{1}{4}\partial_{\mu}\Phi_{2}^-\partial^{\mu}\Phi_{2}^+
-\frac{1}{4}\partial_{\mu}\Phi_{1}^0\partial^{\mu}\Phi_{1}^0
+\frac{ig}{4}A^\mu(\Phi_{2^-}\partial_{\mu}\Phi_{2}^+ - \Phi_{2^+}\partial_{\mu}\Phi_{2}^-)
\end{equation}
\begin{equation}\label{SYM-Lag-higgs}
-\frac{g^2}{2}\Phi_{2^-}\Phi_{2}^+ A_\mu A^\mu
-\frac{g^2}{4}(\Phi_1^0+v)^2\Phi_{2^-}\Phi_{2}^+.
\end{equation}
Here $F_{\mu\nu}=\partial_\mu A_\nu-\partial_\nu A_\mu$ and we have identified $A_\mu^3=A_\mu$. A crucial step here is to implement a constraint 
 \begin{equation}\label{constraint}
 \Phi^0_1+\alpha/r=0,
 \end{equation}
 into the Lagrangian (\ref{SYM-Lag-higgs}).  Bellow we will see why this contraint is crucial in our construction. After the strong implementation of this contraint into the Lagrangian (\ref{SYM-Lag-higgs}) we obtain\footnote{This constraint can be interpreted as a second class constraint in the Dirac sense.}
\begin{equation}\nonumber
{\cal L}= -\frac{1}{8}F^{\mu\nu}F_{\mu\nu}
-\frac{1}{4}\partial_{\mu}\Phi_{2}^-\partial^{\mu}\Phi_{2}^+
+\frac{ig}{4}A^\mu(\Phi_{2^-}\partial_{\mu}\Phi_{2}^+ - \Phi_{2^+}\partial_{\mu}\Phi_{2}^-)
\end{equation}
\begin{equation}\label{SYM-Lag-red}
-\frac{g^2}{2}\Phi_{2^-}\Phi_{2}^+ A_\mu A^\mu
-\frac{g^2}{4}(-\frac{\alpha}{r}+v)^2\Phi_{2^-}\Phi_{2}^+
\end{equation}
up to boundary term. The field content of our theory is reduced to one vector field $A_\mu$ and a complex scalar $\Phi_2$.

The equation of motion for the scalar field $\Phi_2^-$ is
\begin{equation}\label{1111}
\partial_{\mu}\partial^{\mu}\Phi_{2}^{-}-ig(\partial_{\mu}A^{\mu})\Phi_{2}^{-}-2igA_{\mu}\partial^{\mu}\Phi_{2}^{-}-g^{2}A_{\mu}A^{\mu}\Phi_{2}^{-}-g^{2}(v-\frac{\alpha}{r})^{2}\Phi_{2}^{-}=0 \, .
\end{equation}
Of course an analogous equation of motion can be obtained for the complex conjugate $\Phi_2^+$ that we will not need here. Denoting $\Phi^-_2=\phi$ and taking the Coulomb potential $A^{\mu}=(-\alpha/r,\boldsymbol{0})$ with $g=1$,\footnote{Which is equivalent to absorbing the coupling constant into $\alpha$.} the equation (\ref{1111}) becomes
\begin{equation}
\partial_{\mu}\partial^{\mu}\phi-2iA^{\mu}\partial_{\mu}\phi-A^{\mu}A_{\mu}\phi-\left(m-\frac{\alpha}{r}\right)^{2}\phi=0,\label{eq:kgmodrel}
\end{equation}
where the vacuum expectation value of $\Phi_{1}$ is the mass $m$ of the scalar field $\phi$. The scalar field equation (\ref{eq:kgmodrel}) that comes from  ${\cal N}=4$ SYM scalar couplings and the Higgs mechanism with the constraint (\ref{constraint}) just presented will be central to our work. Among the crucial interesting properties of (\ref{eq:kgmodrel}) is that the non-minimal coupling $m\to (m-{\alpha}/{r})$ induced by the Higgs mechanism and the constraint $\Phi^0_1=-\alpha/r$ cancels out the quadratic term coming from the minimal coupling (introduced through the covariant derivative), enhancing the symmetry of the resulting field theory from SO(3) to SO(4). 

The Lagrangian (\ref{SYM-Lag-red}) can be rewritten in terms of the fields $\phi,\phi^*, A_\mu$ as
\begin{equation} \label{FAphi-lagrangian}
{\cal L}=-\frac{1}{4}F^{\mu\nu}F_{\mu\nu}-\frac{1}{2}(D_{\mu}\phi)^{\ast}(D^{\mu}\phi)-\frac{1}{2}\left(m-\frac{\alpha}{r}\right)^{2}\phi^{\ast}\phi,
\end{equation}
and recognize this Lagrangian as the scalar electrodynamics with modified mass. Here the covariant derivative is $D_{\mu}=\partial_{\mu}-iA_{\mu}$. The equations of motion for $A^{\nu}$ are
\begin{equation}\label{eq:eqfora}
-\partial_{\mu}\partial^{\mu}A^{\nu}+\partial^{\nu}(\partial\cdot A)=\frac{i}{2}(\phi\partial^{\nu}\phi^{\ast}-\phi^{\ast}\partial^{\nu}\phi)-A^{\nu}|\phi|^{2},
\end{equation}
The application of the divergence to (\ref{eq:eqfora}) reveals that the conserved current is
\begin{equation}
J^{\nu}=\frac{i}{2}(\phi\partial^{\nu}\phi^{*}-\phi^{*}\partial^{\nu}\phi)-A^{\nu}|\phi|^{2}.
\end{equation}
This means that the density $\rho$ is
\begin{equation}
\rho=\frac{i}{2}(\phi^{*}\partial_{t}\phi-\phi\partial_{t}\phi^{*})+\frac{\alpha}{r}|\phi|^{2},
\end{equation}
taking into account that $A^\mu=(-\frac{\alpha}{r},\bf 0)$. From the other hand, the implementation of this Coulomb potential in the equation of motion (\ref{eq:eqfora}) gives
\begin{equation}
\nabla^{2}\left(\frac{\alpha}{r}\right)=\frac{i}{2}(\phi^{*}\partial_{t}\phi-\phi\partial_{t}\phi^{*})+\frac{\alpha}{r}|\phi|^{2},
\end{equation}
which implies
\begin{equation}
\rho=\nabla^{2}\left(\frac{\alpha}{r}\right)
\end{equation}
as expected, or 
\begin{equation}
\int\rho\mathrm{d}^{3}r=-4\pi\alpha.
\end{equation}
In this way we have checked that our prescriptions are entirely consistent. Even though the mass of the associated scalar electrodynamics is replaced by the non minimal coupling $m\to (m-{\alpha}/{r})$ the associated charge of the scalar field is the expected one.

\section{Relativistic particle and Runge-Lenz vector}\label{sec:relpart}

In this section we will show that the field equation (\ref{eq:kgmodrel}) has an enhanced symmetry SO(4). For that end we start from the construction of the relativistic particle of mass $m$ implementing the standard minimal coupling to a scalar potential followed by a non-minimal prescription suggested by equation (\ref{eq:kgmodrel}). Then we will construct the analogous of the RL vector for this particle. This relativistic RL vector is the generator of the enhanced SO(4) symmetry.  

The standard formulation of a relativistic particle minimally coupled to a background electromagnetic field $A_\mu$ starts from the action
\begin{equation}\label{eq:actbefevt}
S=-mc\int d\tau \sqrt{-\eta_{\mu \nu} \frac{d x^\mu}{d\tau}\frac{d x^\nu}{d\tau}} + \frac{1}{c} \int d\tau A_\mu \frac{dx^\mu}{d\tau},
\end{equation}
where $\eta_{\mu \nu} =\text{diag} \left(-,+,+,+ \right)$ and $\alpha$ is the coupling constant.

The canonical momenta 
\begin{equation}
p_\mu =\frac{\partial L}{\partial \dot{x}^\mu} = \frac{mc \ \eta_{\mu\nu}\dot{x}^\nu}{\sqrt{-\eta_{\alpha \beta} \frac{d x^\alpha}{d\tau}\frac{d x^\beta}{d\tau}}} +\frac{1}{c} A_\mu,
\end{equation}
imply the quadratic constraint (associated to reparametrization invariance)
\begin{equation} \label{eq:constasac}
\eta^{\mu \nu}\left( p_\mu - \frac{1}{c} A_\mu \right) \left( p_\nu - \frac{1}{c} A_\nu \right) + m^2 c^2=0,
\end{equation}
that leads naturally to the Klein-Gordon (KG) equation upon the identification $p_\mu\to -i\hbar\partial_\mu$
\begin{equation}\label{actrp1}
\left[ \eta^{\mu \nu } \left( -i\hbar \partial_\mu  - \frac{1}{c} A_\mu \right) \left(-i\hbar \partial_\nu - \frac{1}{c} A_\nu \right) + m^2c^2 \right]\phi =0.
\end{equation}

In the particular case of a Coulomb field $A_0= \alpha/r, A_i=0$ and working in units such that $\hbar=1, c=1$, the equation above reads
\begin{equation}
 \left[  \left(  -i\partial_t - \frac{\alpha}{r}\right)^2 + \nabla^2 - m^2    \right]\phi(x) =0.
\end{equation}
The corresponding stationary spectrum with energy $E_{\ell,n}$ that comes from this equation is the usual relativistic spectrum of the KG equation with the external Coulomb potential \cite{Sakata,Jackiw}
$$ E_{\ell,n}=m\left[1+\left(\frac{\alpha}{n+\sqrt{(\ell+1/2)^2-\alpha^2}-(\ell+1/2)}\right)\right]^{-1/2},$$
where we can see that the well known degeneracy of the associated  non-relativistic spectrum is broken. Apparently, in the relativistic case, we can not construct the conserved RL vector associated with the above degeneracy of the spectrum.  A crucial observation is that the SO(4) symmetry (of the non-relativistic spectrum) is broken by the quadratic term $\alpha^2/r^2$ that comes from the minimal coupling to the Coulomb potential in the KG equation (\ref{actrp1}).
 
The {\em classical} stationary problem with conserved angular momentum $\ell$ and  energy $E=-p_0$ is
$$(E-V)^2-(p_r^2+\ell^2/r^2)-E_0^2=0,$$
where $E_0\equiv m$. The solutions for bounded orbits are rosettes \cite{Landau} in contrast with the non-relativistic problem where the orbits are ellipses \cite{Goldstein}. As a consequence, the corresponding RL vector is not conserved in the relativistic theory described by the Lagrangian (\ref{eq:actbefevt}). 
 
To see that it is possible to retain the degeneracy of the non-relativistic spectrum in the relativistic case we will follow our motivation from (\ref{eq:kgmodrel}), introducing a non-minimal coupling given by the substitution 
\begin{equation}\label{masst}
m \longrightarrow m-\frac{\alpha}{r}. 
\end{equation}
This new coupling restores the degeneracy of the associated non-relativistic dynamics and as a consequence a {\em new} relativistic symmetry emerges.  The action (\ref{eq:actbefevt}) is now
\begin{equation}\label{act12}
S=\int d \tau \left[ -m \sqrt{- \left( 1-\frac{\alpha}{rm}\right)^2 \eta_{\mu \nu} \dot{x}^\nu \dot{x}^\mu} + \frac{\alpha}{r} \dot{x}^0\right].
\end{equation}
This action corresponds to a relativistic particle interacting with a Coulomb background in a curved space given by a conformally flat metric
$$\eta_{\mu\nu}\to \left( 1-\frac{\alpha}{rm}\right)^2 \eta_{\mu \nu}.\label{conf-metric}$$
The associated  constraint (\ref{eq:constasac}) is now
\begin{equation} \label{eq:constpo}
p_0^2- \frac{2 \alpha}{r} p_0 - \vec{p}^2 - m^2 + \frac{2\alpha m}{r}=0.
\end{equation}

Considering the {\em classical} stationary problem, using the non-minimal replacement (\ref{masst}) to remove the anomaly, and denoting $A_0=-V$ we have 
$$E_0^2\to (E_0+V)^2$$
and the modified KG equation reads
$$E^2-E_0^2-2V(E_0+E)- (p_r^2+\ell^2/r^2)=0$$
or
$$(E+E_0)\left(E-E_0-2V-\frac{1}{(E+E_0)}( p_r^2+\ell^2/r^2)\right)=0,$$
where $\ell$  is the conserved angular momentum. An equivalent way to write this result is
$$(E-E_0)-2V-\frac{1}{(E+E_0)}( p_r^2+\ell^2/r^2)=0,$$
where we observe that the relativistic problem is reduced to a Schr\"odinger like problem. If we want to obtain the associated non-relativistic (NR) problem just replace\footnote{$E=\frac{m}{\sqrt{1-v^2}}$, then in the NR limit $E=E_0+mv^2/2$, or $E-E_0\to E_{NR}$. Also, from $E^2-E_0^2=p^2$ and then $(E+E_0)(E-E_0)=p^2$, we have $E_{NR}=p^2 /(E+E_0)$ and consequently $1/(E+E_0)\to \frac{1}{2m}$.}
$$(E-E_0)\to E_{NR}$$
and
$$\frac{1}{(E+E_0)}\to \frac{1}{2m}.$$
It is also true that starting from the non-relativistic problem we can obtain the relativistic one by reading the same replacements from right to left.  
 
The hydrogen like spectrum that arises from this modified KG equation (\ref{eq:kgmodrel}) recovers the full degeneracy of the non-relativistic case \cite{Sakata}
$$E_n=m\left(1-\frac{2\alpha^2}{n^2+\alpha^2}\right).$$
As a consequence, it is now  possible to construct a relativistic analogue of the RL vector. 

Using as a model the non-relativistic construction \cite{Goldstein} we find
\begin{equation}\label{eq:lrlgen}
\frac{d}{d\tau} \left[ \vec{p}\times\vec{L}+{\alpha} \left( p_0-m\right) \frac{\vec{x}}{r}\right]=0,
\end{equation}
where $\vec L$ is the angular momentum that generates the corresponding SO(3) algebra.     The relativistic  generalization of the Runge-Lenz vector is then
\begin{equation}
\vec{K}= \vec{p}\times\vec{L}-\left( m-{p_0}\right)\frac{\alpha\vec{x}}{r}.
\end{equation}
or
\begin{equation}
\vec{K}= \vec{p}\times\vec{L}-\left( E_0+E\right)\frac{\alpha\vec{x}}{r}\label{RRL}.
\end{equation}
So we recover the conservation of the RL vector and the orbits are closed ellipses. The vector $\vec{K}$ enhances the symmetry from SO(3) to SO(4), and in this way we have shown that the non-minimal coupling induced by (\ref{masst})  or the transformation to a conformally flat space in  (\ref{conf-metric}) allows us to recover the SO(4) symmetry in the relativistic case.

Now, taking the non-relativistic limit of the RL vector we obtain
\begin{equation}
\vec{K}_{NR} = \vec{p} \times \vec{L} - 2\alpha m \frac{\vec{x}}{r},
\end{equation}
which is the usual RL vector with a coupling constant that is twice larger.\\

{\em {\bf Remarks:} \\

{\bf a)} A curious feature about the non-minimal coupling (\ref{masst}) is that the NR effective potential has a factor of 2 as compared with  the usual NR formulation.\\

{\bf b)} To every NR (relativistic) observable  we can associate a relativistic (NR) observable with the replacement $2m \leftrightarrow E+E_0$, $\,\,  E_{NR}\leftrightarrow E-E_0$.\\

{\bf c)} Notice that the substitution (\ref{masst}) has a critical point  when $r_c\equiv \frac{\alpha}{m}$.  At this point the mass term in the Lagrangian (\ref{FAphi-lagrangian}) is zero. The kinetic term of the particle Lagrangian (\ref{act12}) is also zero because the conformal factor of the Minkowski metric tends to zero. This can be contrasted with the behaviour of a relativistic particle in the limit $m\to 0$. Hence, starting with the equivalent description of the Lagrangian (\ref{act12}) in terms of the einbein $e$
\begin{equation} 
S=\int d\tau \left( \frac{\dot{x}_\mu \dot{x}^\mu}{2e} - \frac{e}{2}\left(m-\frac{\alpha}{r} \right)^2 + \frac{\alpha}{r} \dot{x}^0\right),
\end{equation}
we can see that in the limit $r\to r_c$ the only term that remains is
\begin{equation} 
S=\int d\tau \left( \frac{\dot{x}_\mu \dot{x}^\mu}{2e}+ m \dot{x}^0 \right),
\end{equation}
since the last term is just a total derivative, reflecting the fact that $p_0$ is defined up to a constant shift $p_0\to  m-\dot x^0/e$. This theory is consistent with (\ref{eq:kgmodrel})  but with $A^\mu=(-m,\bf 0)$, corresponding to a KG equation with zero mass term minimally coupled to a constant potential.  \\

{\bf d)} A central argument given here is that the unusual coupling (\ref{masst}) comes from the scalar sector of ${\cal N}=4$ SYM theory with an appropriately adjusted Higgs mechanism as presented in Section 2. This construction is based on a powerful conformal field theory which has very interesting integrability properties. Due to the hidden symmetry that lies under the replacement (\ref{masst}) that reveals the existence of the relativistic RL vector (\ref{RRL}), the ${\cal N}=4$ SYM theory is sometimes dubbed as the hydrogen atom quantum field theory \cite{bruser-caron-henn}.}

\section{The relativistic SO(4) algebra using the relativistic RL vector}

The aim of this section is twofold. On one hand we will construct the infinitesimal Noether symmetries generated by the relativistic RL vector and observe that this symmetry is not an SO(4) rotation but it is neither a conformal symmetry. A deeper analysis is needed to compare the symmetry obtained here with the symmetry in the dual momentum space presented in \cite{Caron}. As a spinoff we show that the relativistic orbit can be reconstructed using the relativistic RL vector.

On the other hand we will describe the complete SO(4) algebra generated by the angular momentum and the new relativistic RL vector. We will restrict ourselves to the stationary problem with relativistic energy ${E}$. In a second step we will recover from this algebra the correct relativistic spectrum of the corresponding hydrogen-like atom in the two body bound state.

From equation (\ref{eq:constpo}) and  solving for $p_0$ 
\begin{equation}
p_0=-\frac{\alpha}{r} \pm \sqrt{\frac{\alpha^2}{r^2} + \vec{p}^2 + m^2-\frac{2m\alpha}{r}},
\end{equation}
we construct the energy function or equivalently the associated relativistic Hamiltonian 
\begin{equation}
H= -  p_0=E.
\end{equation}
 In this form, we confirm that $p_0$ is a constant of motion and in consequence invariant under the transformations generated by the Runge-Lenz vector,
\begin{equation}
\delta p_0 = \left\lbrace  p_0 , \epsilon_i K^i  \right\rbrace =0,
\end{equation}
The infinitesimal transformations associated with $r^i$ and the canonical momenta $p_i$ are (setting again $c=1$) 
$$\delta r^i=\left\lbrace r^i, \epsilon_j K^j \right\rbrace=2(\epsilon\cdot r)p_i-r^i (\epsilon\cdot p) -(r\cdot p)\epsilon^i ,
$$
$$
\delta p_i = \left\lbrace p_i, \epsilon_j K^j \right\rbrace = - (\vec{p})^2 \epsilon_i +(\vec{p} \cdot \vec{\epsilon})p_i + \alpha\left({p_0} -m \right) \frac{\vec{\epsilon} \cdot \vec{x}}{r^3} x_i- \alpha\left({p_0} -m \right)\frac{\epsilon_i}{r},
 $$
 or
$$ \delta p_i= - (\vec{p})^2 \epsilon_i +(\vec{p} \cdot \vec{\epsilon})p_i - \alpha\left(E +E_0 \right) (\frac{\vec{\epsilon} \cdot \vec{x}}{r^3} x_i- \frac{\epsilon_i}{r}).
$$
The Runge-Lenz vector also acts on the magnitud $r$ as
 \begin{equation}
 \delta r = \left\lbrace \sqrt{x_l x_l}, \epsilon_i K^i \right\rbrace = \frac{\left( \vec{p}\cdot \vec{x} \right) \left(\vec{x}\cdot \vec{\epsilon} \right)}{r} - (\vec{\epsilon} \cdot \vec{p}) r, 
 \end{equation}
These infinitesimal symmetries do not correspond exactly with the symmetry transformations previously written in \cite{Caron}. The reason for the mismatch and its possible consequences is an open problem that we will leave for future work. A crucial difference is that in the approach given in \cite{Caron}  the symmetry transformation acts in a dual momentum space (dual conformal transformation)  that is appropriate to reveal the symmetries of scattering amplitudes in SYM theory.

We observe that the analogous procedure to obtain the classical non relativistic orbit of the Kepler problem can be implemented also in the relativistic case.
If we take the dot product of the RL vector with $\vec{x}$, we obtain
\begin{equation}
\vec{K} \cdot \vec{x}=Kr \cos\theta = \left(\vec{p}\times\vec{L}\right)\cdot \vec{x}-\alpha \left( E_0+E\right) \frac{\vec{x}\cdot \vec{x}}{r},
\end{equation}
or
\begin{equation}
\frac{L^2}{r \left(E_0+E \right)\alpha} = \frac{K \cos \theta}{\alpha \left( E_0+E\right)}+1,
\end{equation}
that is, the equation of a conic with eccentricity
\begin{equation}
\varepsilon=\frac{K}{\alpha \left( E_0+E\right)}.
\end{equation}
Here, we notice that under the substitution of $(E_0+E)\to 2m $ we recover the NR result of \cite{Goldstein}. 

The complete spectrum can also be constructed from the relativistic SO(4) algebra generalizing the NR result as presented in \cite{gilmore, Paddy}. Let us start from a simple redefinition of the relativistic RL vector (\ref{RRL}),
$$  {\vec A}=\frac{2}{E+m}\vec K=\frac{1}{{E}+m} (\vec p\times \vec L-\vec L\times \vec p)-2\alpha \frac{\vec r}{r},$$
for a bounded orbit with energy $E$. We know from our previous calculation that
$$[\vec A,H]=0,\qquad\vec  L\cdot \vec A=\vec A\cdot\vec L,$$
and 
\begin{equation}\label{EqK2}
A^2=4\left(\alpha^2+\frac{{E}-m}{{E+m}}(1+L^2)\right).\end{equation}
The corresponding relativistic algebra closes as
$$[L_i,L_j]=i\varepsilon_{ijk}L_k,$$
$$[A_i,L_j]=i\varepsilon_{ijk}A_k,$$
$$[A_i,A_j]=-4i\left(\frac{E-m}{E+m}\right)\varepsilon_{ijk}L_k,$$

Defining
$${\vec A}'=\sqrt{-\frac{E+m}{4(E-m)}}\vec A$$
and
$$\vec N=\frac12(\vec L+{\vec A}'), \qquad \vec M=\frac12(\vec L-{\vec A}')$$
it is easy to show that the original algebra splits into the product of two SO(3) algebras,
$$[N_i,N_j]=i\varepsilon_{ijk}N_k,\qquad [M_i,M_j]=i\varepsilon_{ijk}M_k,$$
with the constraint
\begin{equation}N^2=M^2\label{NM-C}.\end{equation}
The operator
$$N^2+M^2$$
will have the eigenvalues $2\ell(\ell+1)$ with $\ell=0,1,2\ldots$ (2 times the square of the angular momentum eigenvalues)  because of the constraint (\ref{NM-C}). On the other hand
$$\frac12(N^2+M^2)=\frac12 [L^2-\frac{E+m}{4(E-m)} A^2]=-\frac{E+m}{2(E-m)}\alpha^2-\frac12 ,$$
where we used (\ref{EqK2}). So, we obtain for the relativistic spectrum
\begin{equation}\label{jap-spectra}
n^2=-\frac{E+m}{E-m}\alpha^2 .
\end{equation}
This is the correct relativistic spectrum reported in \cite{Sakata} and reproduced here with $n=2\ell+1$.

\section{Relativistic KS duality}

The KS  dictionary can be constructed from the NR case (see for example \cite{CG}). Starting from a potential of the form $V=k r^\beta$ and introducing a change of variable $r\to R^{\frac{2}{\beta+2}}$, the integral for the orbit \cite{Goldstein} for the variable $R$ has the same form as the original orbit integral for $r$ if we define  
$$
{\cal V}=-(E-E_0)R^{-\frac{2\beta}{\beta+2}}$$
as the new potential, and the new energy by
$${\cal E}-{\cal E}_0=-2k.$$
A crucial difference from the NR case is the factor 2 in the new energy definition. Also the new angle of the orbit in the relation $r(\theta)\to R(\Theta) $ must be rescaled by a factor: $\Theta=((\beta+2)/2)\theta$.

\st{Just}As a consistency condition we also need the identification
$$E+E_0\to {\cal E}+{\cal E}_0$$
that is not present in the NR limit because the original NR KS transformation by construction, maps problems with the same mass. We note that $E-E_0$ and $E+E_0$ play very different roles in the relativistic KS mapping. While $E-E_0$ is a coupling constant in the new problem, $E+E_0$ plays the role of the mass parameter. Of course, we are restricting the mapping in such a way that the rest energies are equal (because the mass of the new and old problems is the same as in the NR KS transformation).  

The new problem is then
$$({\cal E}-{\cal E}_0)-2{\cal V}(R)-\frac{1}{({\cal E}+{\cal E}_0)}( p_R^2+\ell^2/R^2)=0$$
A simple comparison with the original problem
$$(E-E_0)-2V-\frac{1}{(E+E_0)}( p_r^2+\ell^2/r^2)=0$$
reveal that the structures of the old and the new problems are exactly the same. The crucial difference between this equation and the original one worked out in \cite{Landau} is the non-minimal coupling (\ref{masst}).

\section{KS and the  relativistic equivalent Schr\"odinger equation (RSE)}

Accordingly with the previous section,  the radial part of the relativistic Schr\"odinger equation (RSE) in $d$ dimensions is
\begin{equation}\label{S1}
\left(-\frac{1}{(E+m)}\left(\frac{d^2}{dr^2}+\frac{(d-1)}{r}\frac{d}{dr}-\frac{\ell(\ell+d-2)}{r^2}\right) +2V(r)-(E-m)\right)R(r)=0.
\end{equation}
$V(r)$ is a central potential defined as $V(r)=k r^\beta$.
By the change of variable
\begin{equation}
r= \rho^{2/(\beta+2)},
\end{equation}
the Schr\"odinger equation becomes
\begin{equation}
\Bigg(-\frac{1}{(E+m)}\left(\frac{d^2}{d\rho^2}+\frac{(\beta+2(d-1))}{(\beta+2)\rho}\frac{d}{d\rho}-\frac{\ell(\ell+d-2)}{\rho^2}\left(\frac{2}{\beta+2}\right)^2\right) 
\end{equation}
$$-\left(\frac{2}{\beta+2}\right)^2\rho^{\frac{-2\beta}{\beta+2}}(E-m)+2\left(\frac{2}{\beta+2}\right)^2  K\Bigg)\tilde R(\rho)=0,
$$
where  $\tilde R(\rho)=R(r(\rho))$. Now  use the following dictionary (duality) to associate the relevant quantities that define the RSE, $E,K,d,\ell$ to a new set of quantities that parametrize the new system ${\cal E}, \cal{V},D,{\cal L}$.

The dimension\footnote{The case $d=2$ (conformal point) deserves special attention. See \cite{hoj} for details.} maps to a new dimension $D$
\begin{equation}\label{D}
D=\frac{2(\beta+d)}{\beta+2}.
\end{equation}
The energy ${\cal E}$ of the new stationary RSE is related with the coupling constant of the old potential $V=Kr^\beta$ by
\begin{equation}
{\cal E}-m= -2(\frac{2}{\beta+2})^2 K.
\end{equation}
In the same way the new angular momentum ${\cal L}$\footnote{Here we are restricting ourselves to the case of integer new dimension $D$ and integer new momenta ${\cal L}$.} is related with the old angular momentum $\ell$ by 
\begin{equation}
{\cal L}= \frac{2}{\beta+2} \ell .
\end{equation}
Finally the new potential ${\cal V}$ is related with the old energy $E$ by
\begin{equation}\label{V}
{\cal V}= - \frac12 (E-m) \left(\frac{2}{\beta+2}\right)^2\rho^{-\frac{2\beta}{\beta+2}}.
\end{equation}
A crucial observation is  the identification
\begin{equation}\label{Id}
E+m\to {\cal E}+m,
\end{equation}
that follow as a consistency condition.
Using this dictionary the new RSE 
\begin{equation}
\left(-\frac{1}{{\cal E}+m}\left(\frac{d^2}{d\rho^2}+\frac{(D-1)}{\rho}\frac{d}{d\rho}-\frac{\cal{L}(\cal{L}+D-2)}{\rho^2}\right) +2{\cal V}(\rho)-({\cal E}-m)\right)\tilde R(\rho)=0
\end{equation}
acquires the {\em same} form as the original one but with dimension $D$, angular momentum ${\cal L}$, energy ${\cal E}$ and potential ${\cal V}(\rho)$ given by (\ref{D}-\ref{V}) and the identification (\ref{Id}).

Notice that we can not map {\em every} solution of the old stationary problem into a stationary solution of the new problem by this duality. We can only map every {\em bounded}  stationary solution of the old problem into a bounded stationary solution of the new problem. 

We will use this dictionary for the case of the hydrogen atom $\beta=-1$ in $d=3$. In that case, the new problem is the isotropic harmonic oscillator in $D=4$ and angular momenta ${\cal L}=2\ell$.

The energy spectrum of the relativistic hydrogen atom is \cite{Sakata, Qiang}
$$ (2n+2\ell +2)\sqrt{m-E}-2\alpha\sqrt{E+m}=0,$$
and from here we have
$$E=m(1-\frac{2\alpha^2}{N^2+\alpha^2}),$$
with $N=n+\ell+1$.
According to our dictionary the new potential is
$${\cal V}=-2(E-m) \rho^2,$$
so the new coupling constant of the corresponding oscillator is $k=2(E-m)$.

On the other hand, the energy spectrum of the relativistic oscillator in $D=4$ {arising from the dictionary reads} 
$${\cal E}-m=2(4n+2{\cal L}+4)\sqrt{\frac{E-m}{{\cal E}+m}}.$$
This is the energy spectrum of a $D=4$ relativistic oscillator with potential ${\cal V}$. It can be compared with the result given in \cite{Qiang2}
$${\cal E}-m=2(4n+2{\cal L}+4)\sqrt{\frac{k}{2({\cal E}+m)}}, $$
for a particle of mass $m$ in a potential $V=kr^2$.
This matches precisely with our result. So we conclude that the relativistic KS transformation relates the relativistic harmonic oscillator in 4 dimensions with the relativistic Coulomb potential in 3 dimensions. This example shows the power of the relativistic KS transformation constructed here to relate different potentials.

\section{Relativistic spectrum from cusp anomalous dimension}

An interesting calculation of the non-relativistic spectrum of the hydrogen atom from perturbation theory in ${\cal N}=4$ SYM was presented in \cite{Caron}. 
The starting point is SYM and the relativistic symmetry associated with the RL vector.  Nevertheless, the explicit computation of the binding energy  spectrum of the bound state results in the {\em non-relativistic} well-known formula for the hydrogen-like spectrum. Here we will extend the result presented in \cite{Caron} to the full relativistic case. 

Our starting point is an enhanced formula for the energy of the bound state in terms of the cusp angle. Taking into account only the binding energy we define
$$(E_n^b-m)= (E_n^b+m) (\sin\frac{\phi_n}{2}-1),$$
where $E_n^b$ is the relativistic binding energy of the bound state and $\phi_n$ the corresponding cusp (scattering) angle. The quantization condition is 
$$\Gamma_{cusp}(\phi_n)=-n,$$
where $\Gamma_{cusp}$ is the cusp anomalous dimension and $n$ and integer. 

From the other hand, $\Gamma_{cusp}$ has been computed  for weak 't Hooft coupling $\lambda<<1$  and the result is
$$\Gamma_{cusp}({\phi_n})=-\frac{\lambda}{8\pi^2}{\phi_n}\tan \frac{\phi_n}{2}.$$
Since $\lambda$ is small the scattering is small $\phi\approx \pi-\delta$, with
$$\delta\approx \frac{\lambda}{4\pi n}.$$
The solution for the binding energy is
\begin{equation}\label{Eb}
(E_n^b-m)=\frac{\delta^2}{8}(E_n^b+m)=\frac{\lambda^2}{128\pi^2n^2}(E_n^b+m).
\end{equation}
The full relativistic spectrum is
$$E-2m=E^b,$$
where we have subtracted the threshold energy $2m$ and $E_b$ is given by (\ref{Eb}). This result matches the computed relativistic spectrum given in \cite{Sakata} and reproduced here using different approaches. Notice that the dependence in $\lambda^2$ is consistent with the expectation that $\sqrt\lambda\sim e$ \cite{Correa2,Fiol} and the fact that the hydrogen energy spectrum goes like $e^4$. 

We remark that the computation presented in \cite{Caron} is entirely consistent. In the weak coupling (small $\lambda$) and large angular momentum ($n>>1$) the approximation to leading order for the tiny effect $E-2m$  is the NR spectrum. 
 We have presented here the perturbative (still weak coupling) relativistic spectrum for the total binding energy. This result could be confirmed by the $E-J$ Chew-Frautschi plot for different values of $\lambda$ in the weak coupling case ($J>>1, E\sim 2m$). 
 
 As a final comment notice the interesting relation between the stereographic projection (small circles on the sphere to large circles on the plane) related by the RL symmetry (as presented for example in \cite{gilmore}), and the duality between static quarks on $S^3\times R$ and dynamical quarks in the plane \cite{Correa, Caron1} as compared with the classical duality between the free particle in $S^3$ with the Kepler problem in $R^4$ \cite{gilmore}. This observation needs further analysis that we will not address here.

\section{Acknowledgment}
This work was supported in part by DGAPA-PAPIIT grant IN103716, CONACyT project 237503 and scholarship 419420 (J.A.J.).  
JAG was partially  supported by Mexico National Council of Science and Technology (CONACyT) grant 238734 and DGAPA-UNAM grant IN107115.\\

\end{document}